# Challenging Data Aggregation Practices: A MAIHDA Study of Asian Student Outcomes in Introductory Physics

Vy Le[1], Grace Angell[1], Jayson Nissen[2], Ben Van Dusen[1]
[1]*School of Education, Iowa State University, Ames, Iowa 50011, USA*
[2]*Department of Physics, Montana State University, Bozeman, Montana 59715, USA*

Acknowledgement
This work is funded in part by NSF Grants No. 2322015 and 2141847. We are grateful to LASSO for their support in both collecting and sharing data for this research.

Conflict of interest statement:
The authors declare no competing financial or personal interests that could have influenced the work reported in this paper.

## Abstract

Aggregation of Asian student data can reinforce the model minority myth by obscuring educational disparities among Asian student subgroups. This study investigated variation in conceptual physics knowledge across Asian racial and ethnic subgroups using data from the LASSO platform, analyzing responses from 16,810 students enrolled in 493 introductory calculus-based physics courses across 64 U.S. institutions. We applied Multilevel Analysis of Individual Heterogeneity and Discriminatory Accuracy to examine predicted pre- and posttest performance on the Force Concept Inventory and Force and Motion Conceptual Evaluation. The findings revealed performance differences among 19 Asian subgroups that the pan-Asian strata (the single aggregated Asian group) concealed. Subgroup predicted means spanned 15.8 percentage points on the pretest and 15.4 percentage points on the posttest. The lowest-performing subgroup's posttest mean was roughly equal to the highest-performing subgroup's pretest mean, indicating a performance gap of about a full semester of instruction. Mean absolute error between the pan-Asian strata and the 19-subgroup estimates was 3.9 percentage points at pretest and 4.0 percentage points at posttest, equivalent to approximately 4-5 weeks of learning in a 16-week course. These findings demonstrate that fine-grained identity data collection can support identifying disparities that common aggregation practices conceal.

## Introduction

The model minority myth is a stereotype that homogenizes all Asian (Americans) into a monolithic group of overachievers (Museus & Kiang, 2009; Wu, 2013; Zhang et al., 2025). This stereotype often obscures and renders invisible the diverse experiences of Asian (American) communities. Researchers have documented how policymakers, media, and dominant institutions weaponize the “model minority” stereotype of Asian (Americans), portraying them as high-performing despite their minority status, to undermine efforts toward racial equity and to blame Black Americans for systemic disparities (C. J. Kim, 1999; Museus & Kiang, 2009; Poon et al., 2016; Wu, 2013). In educational research, the model minority myth often contributes to the aggregation of all Asian (American) students into a single category. Several works called for disaggregating the broad Asian categorization in physics and STEM fields, as this can reveal disparities masked by the model minority myth (Li & Zhao, 2025; Sbeglia & Nehm, 2024; Shafer et al., 2021; Zhang et al., 2025).

## Research question

In this article, we examine the common assumption that Asian (Americans) are a homogenous, high-achieving group. Through this, we challenge the practice of aggregating Asian students into a single group or with White students without being aware of and transparent about the impacts of these choices on model outcomes and conclusions.

Disaggregated data is necessary, though not sufficient, for developing and implementing educational policies and practices that serve all students and broaden participation in physics and STEM education.

- To what extent does average physics conceptual knowledge vary across Asian races/ethnicities at the start and end of introductory college calculus-based physics courses?

## Positionality

Author 1. I identify as an Asian cisgender woman, raised in rural southeast Vietnam. My educational and professional path, from earning a physics degree in Vietnam to pursuing graduate studies in education in the U.S., has spanned transnational and intersectional contexts. I began teaching high school physics in Vietnam and later taught college-level physics in the U.S., working closely with many Asian American students. Cross-cultural experiences with people of different races have enhanced my understanding of educational systems and learners' varied experiences. I advocate for creating inclusive learning environments where students from all backgrounds feel respected, supported, and empowered to succeed.

Author 2. I identify as a White, cisgender woman, and recognize that this, along with my limited physics background, shapes the perspective I bring to the paper. I graduated with a bachelor's degree in psychology and am currently pursuing a master's degree in teaching, where I hope to work with students from rural, low socioeconomic backgrounds similar to the one I grew up in. In my classroom, I hope to practice culturally responsive teaching and focus on building positive relationships with my students. The work I have done with foster youth has informed my awareness of how systemic barriers and limited support can affect educational outcomes and attainment. I am passionate about students recognizing the value and importance of their unique voices and perspectives, and the strengths they bring to academic spaces.

Author 3. As a White man, I recognize that I benefit from systems of privilege and power that shape opportunities in American society, including in science and education. My motivation to pursue this work also comes from my own experiences of how science has been empowering and liberating in my life, and from my desire to make those opportunities more widely accessible. I seek collaborations and projects that push forward my own growth while advancing collective efforts to support students who are marginalized by racism, sexism, classism, and other systems of oppression to pursue their goals and curiosity. This study fits within my work using disaggregated data and intersectional research methods to better measure student outcomes and to inform policies and practices that expand participation and success in STEM.

Author 4. I identify as a White cisgender man. I earned a bachelor's degree in physics, engaged in physics research, was a high school physics teacher, and now have a Ph.D. in education and prepare future science teachers. Raised in low-income households, I now earn an upper-middle-class income. Throughout my career, I have actively engaged with diverse educational contexts, including hosting Japanese exchange students and facilitating student exchanges to Japan. This involvement has afforded me an appreciation of cultural and educational dynamics, particularly in East Asian contexts, though I acknowledge that these experiences do not fully bridge the gap to the lived realities of marginalized communities in the U.S. People with similar privileges to mine have created and maintained our society's unjust power structures. As a person with privilege, I believe it is my obligation to use that privilege to dismantle oppressive systems, while also recognizing that my perspective is limited by that very privilege.

## Definition

To support readers' interpretation of our study, Table I includes a selection of terms and their definitions.

| **Terms** | **Definitions** |
|---|---|
| Heterogeneity | (Statistical) Heterogeneity refers to differences or variations within a population, group, or sample (Schudde, 2018). In this study, heterogeneity is reflected in differences in posterior estimates across racial and ethnic subgroups (e.g., Asian Indian, Chinese, Korean, etc.). |
| Asian (American) | A term refers to all individuals of Asian ancestry residing in the United States, whether temporarily or permanently, regardless of citizenship status (Chen & Buell, 2018; Zhang et al., 2025). |
| Model Minority Myth | A stereotype that homogenizes Asian (Americans) as highly successful, flattening the variety of cultures, languages, and histories among different ethnic groups within this strata (S. J. Lee, 1994; Poon et al., 2016). |
| Critical Race Theory (CRT) | A framework originating in legal studies, created to address racial injustices and oppression. Assumes racism is embedded within institutions, and prioritizes amplifying the narratives of the oppressed (Ladson-Billings & Tate, 1995). |
| Asian Critical (AsianCrit) theory | An extension of critical race theory (CRT) that centers Asian American experiences (Museus & Itikar, 2013). |
| Quantitative Critical Race theory (QuantCrit) | Applies the principles of CRT to quantitative research, analyzing how data and statistical methods can reinforce racial biases and contribute to systemic oppression. Critiques the assumption of neutrality in numbers, advocating instead for using numbers to challenge racial injustices (Gillborn et al., 2018). |
| Intersectional Identities | A term referring to how social identities, such as race, ethnicity, and gender, interact to shape individual experience. The focus of intersectionality is how these identities interact dynamically with power structures, leading to inequities (Crenshaw, 1991). |
| Multiple Imputation (MI) | A statistical approach replaces missing values in the existing data with several sets of plausible estimates that reflect the uncertainty in the missing data. (Rubin, 1987; Woods et al., 2024). |
| Multilevel Analysis of Individual Heterogeneity and Discriminant Analysis (MAIHDA) | A statistical modeling method designed for intersectional research, which leverages multilevel modeling to nest individuals within social strata (Evans et al., 2024). |
| Social Strata | Represent the intersections of social identities such as race, class, and gender within a set of power structures (Evans et al., 2024). For example, “Asian Indian" and "White Korean" are strata. In MAIHDA, researchers use strata to quantify differences in outcomes across these intersections. |

**Table I: Definitions of terms.**

## Literature review

In this section, we begin by tracing the historical development of the model minority myth, which has long portrayed Asian (Americans) as a monolithic and uniformly successful group (Chou, 2016). We then review works that challenge this narrative by documenting variation in educational, economic, and social outcomes across Asian subgroups. Drawing on both historical and contemporary literature, we acknowledged differences in immigration histories, migration motivations (Ogbu & Simons, 1998), and socioeconomic conditions (AAPI Data, 2022) that shape these varied experiences. This literature provides the context for our research question regarding whether disparities in academic performance exist across Asian (American) subgroups.

### Model Minority Myth

The model minority myth and stereotype portray Asian (Americans) as a monolithic group of overachievers (Poon et al., 2016; Zhang et al., 2025). Politicians and journalists have historically used this myth to contrast Asian (Americans) with other marginalized groups, particularly Black Americans, to undermine demands for racial justice and reinforce claims that systemic racism can be overcome through individual effort (Museus & Kiang, 2009). Media narratives often attribute Asian (Americans) success to cultural values, such as a strong work ethic or a focus on education, overlooking the wide-ranging differences in culture, language, and immigration patterns among the more than 40 Asian ethnic groups in the United States (Krogstad & Im, 2025; S. Lee et al., 2017). This myth dismisses the challenges Asian (Americans) have faced and perpetuates the stereotype that all Asian (Americans) are successful. This excludes them from racial discourse and limits research and policy work being done to support them (Museus & Itikar, 2013).

Historical roots of the model minority myth emerged during World War II (Wallace, 2021; Wu, 2013). After China became an ally in World War II, Congress repealed the Chinese Exclusion Act, and the United States began to develop divergent narratives between groups that were once conglomerated. Chinese Americans became allies, and Chinese-born immigrants were newly permitted to become U.S. citizens. Japanese Americans, in contrast, were branded as enemies and were mass imprisoned in concentration or "relocation" camps.

Following World War II, and the 1944 release of Gunnar Myrdal's *An American Dilemma,* social scrutiny of racial relations and discrimination in the U.S. was on the rise (Wallace, 2021; Wu, 2013). Myrdal attributed the racial disparities faced by Black people in America to White prejudice rather than inherent racial differences (Myrdal & Bok, 1996). Many media outlets, e.g., The New York Times, released articles comparing Black Americans to Asian (Americans) to focus the blame for racist disparities on Black Americans. They developed narratives portraying Asian (Americans) as more diligent and hardworking than Black Americans, and leveraged these narratives against Black communities. William Peterson's "Success story: Japanese American style" came out during this time, often attributed to the first articulation of the model minority myth, although the term itself is not used (C. J. Kim, 1999). In his New York Times article, Peterson argues that Japanese Americans succeeded in spite of their own disparities, using their success against Black Americans, who he refers to as "problem minorities" (Petersen, 1966).

The 1965 Watts Riots, a protest against police violence in Black communities in California, are another example that brought the model minority myth into mainstream discourse (Iftikar & Museus, 2018). Daniel Moynihan, the Secretary of Labor, published a report contrasting the high unemployment rates of Black men with the perceived diligence of Asian American families. This reinforced the model minority myth, framing it as though Black communities were at fault for inequalities and triangulating Asian (Americans) as a buffer, "better than Blacks but inferior to Whites" (Iftikar & Museus, 2018; C. J. Kim, 1999).

Shafer et al. (2021) examined course grades in university-level physics courses and found no evidence of racial inequities when using URM and non-URM groupings. However, after disaggregating the data by specific racial/ethnic groups, Shafer et al. revealed that Black and Asian American students underperformed relative to their non-Hispanic White peers, inequities that the URM's aggregated identities had concealed.

When equities are disguised by how data is categorized, groups that could benefit from intervention do not receive it. This relates to the work of Iftikar and Museus (2018), who advocate for the use of more specific frameworks, such as AsianCrit. They suggest that frameworks like AsianCrit can help investigate how systemic racism within educational systems impacts Asian American students. Research that centers the experiences and histories of communities of color can help challenge their marginalization in institutional decision-making and foster a more equitable distribution of resources and outcomes.

## Historical Trends

Historically, Asian (Americans) have been grouped under one broad category, disguising the variety of political, economic, and educational experiences within the Asian categorization (Krogstad & Im, 2025). Poon et al. (2016) suggest that analyzing the educational data of Asian American students should be grounded in a historical context. A report compiled by AAPI Data (AAPI Data, 2022; see Table II) reveals contrasts in educational attainment and income across ethnic groups. Median household income and bachelor's degree attainment range from $125,319 and 75% for Indian Americans to $66,406 and 19% for Native Hawaiians and Pacific Islanders (NHPI) (AAPI Data, 2022; see Table II).

*Asian Indian*

To better understand these differences, we need to consider the historical context of Asian immigration to the United States. The first wave of Indian immigration to America resulted from Britain's colonization and land tenure in India, as many Punjab farmers navigated drought and famine (Rangaswamy, 2008). Between 1903 and 1920, Punjab villagers landed first in Canada, and then migrated down the West Coast of the United States, working primarily for lumber mills and railway companies. Immigration came to a halt with the Barred Zone Act of 1917, a result of race-based hostility that pervaded Indian-immigrants' experience in America during this time (Hess, 1974). The 1965 Immigration Reform Act saw a drastic increase in Indian immigration to the U.S. with the transition of nation-based quotas to hemisphere-based quotas. Most Indian-immigrants travelling to the U.S. were now classified as professional or skilled workers, a contrast to the Punjab villagers from the early 1900s. Today, 54% list employment as their primary reason for migrating (AAPI Data, 2022; see Table II).

*Filipino*

Filipino immigration to the United States has been shaped largely by U.S. military occupation and the annexation of the Philippines (Gates, 1984). The first major wave of Filipino immigration to the United States came following the Philippine-American War. After the Philippines became a U.S. territory, Filipinos were classified as U.S. nationals, allowing them to immigrate to the United States freely (David & Nadal, 2013). During this time, Filipino immigration was concentrated primarily in Hawaii and the western coast of the United States, and consisted mainly of male laborers working for sugar plantations in Hawaii, farms in California, or salmon canneries in Washington and Alaska (Gallardo & Batalova, 2020; Melendy, 1974). Another shift occurred when the Philippines gained independence in 1946, and with the change in immigration legislation in 1965. This time period consisted of a broader variety of Filipino immigrants, including the wives and children of Filipino men in the U.S., military members, students, and professionals ("Immigration History," 2014). Today, half of Filipinos in the United States have a bachelor's degree or higher, and 78% list family as their primary reason for immigration (AAPI Data, 2022; see Table II).

*Southeast Asian*

Southeast Asian groups, such as Cambodian, Lao, Hmong, and Vietnamese Americans, have faced histories of war, displacement, and refugee resettlement. Many arrived in the U.S. as refugees following the Vietnam War, the Cambodian Genocide, and the Secret War in Laos. In April of 1975, President Ford authorized the resettlement of 130,000 refugees from Indochina, most of whom were from Vietnam, due to the imminent fall of Saigon (Bankston & Zhou, 2021; Batalova, 2023; Bon Tempo & Diner, 2022). While the initial wave of Vietnamese immigration to the United States was through refugee status, modern immigration stems primarily through family reunification (Batalova, 2023). As of 2022, 93% of Vietnamese Americans list family as their primary reason for immigration, and this group reported among the highest rates of limited English proficiency (48.9%), and the lowest rates of educational attainment, with 32% receiving a bachelor's degree or higher (AAPI Data, 2022; see Table II).

*Native Hawaiians and Pacific Islanders (NHPI)*

The first Native Hawaiians who migrated to the United States did so between the late 18th and early 19th centuries, brought them to the Northwestern mainland by British fur traders in Hawaii (Quimby, 1972). Hawaii was first settled by Polynesians traveling from the Marquesas Islands between 400-750 CE (Hamilton, 2025; Heckathorn et al., 2025). The arrival of British Captain James Cook in 1778 marked the first European contact with Native Hawaiians. This contact, and the ensuing immigration of Protestant missionaries from New England in 1820, introduced foreign diseases to the island that had a profound impact on the population. Over time, the Native Hawaiian population became a minority in Hawaii, with laborers from other countries such as China, Japan, and the Philippines traveling to Hawaii to pursue work on the sugar plantations, which were owned primarily by immigrants from the mainland United States and Europe. In 1893, the Western plantation owners and missionary descendants overthrew the Hawaiian government, leading to the United States' annexation of Hawaii in 1898 (Hamilton, 2025; Kauana, 2021). Annexation had lasting impacts on the culture, language, health, and homeownership of Native Hawaiians (Kauana, 2021; Pagud et al., 2022). As of 2020, only 47% of Native Hawaiians lived in Hawaii, while 53% lived in the mainland United States (Hamilton, 2025; Pagud et al., 2022). Among Native Hawaiians migrating to the continental U.S., many cited a lack of economic opportunity and affordable housing as contributing factors (Pagud et al., 2022).

Native Hawaiians are the largest group of Pacific Islanders living in the United States today, followed by Samoans and Chamorros, an indigenous group from Guam and the Northern Mariana Islands (Rico et al., 2023). The first major wave of non-Hawaiian Pacific Islanders immigrated to the U.S. in the 1950s, and this group was largely made up of American Samoans and Guamanians. American Samoans became U.S. nationals after American Samoa became an official United States territory in 1900 (U.S. Department of the Interior, 2024). U.S. Naval occupation of American Samoa significantly influenced the first wave of immigration, with many Samoan military personnel immigrating to naval bases in Hawaii after the U.S. Navy withdrew from American Samoa in 1951 (Forster, 1957). The United States annexed Guam in 1898 as a result of the Spanish-American War (Hamilton, 2025). Over 50 years passed before Guamanians were granted full United States citizenship through the Guam Organic Act of 1950 (U.S. Department of the Interior, 2024). Similarly to American Samoans and Filipinos, Chamorro immigration to the mainland United States was influenced by the U.S. Navy occupation, the Korean War, and Typhoon Karen (*Civil Rights Digest*, 1976). Today, the average median salary of NHPI is $66,406, and these groups experience the lowest level of educational attainment, with 19% earning a bachelor's degree or higher (AAPI Data, 2022; see Table II).

*East Asian*

Immigration patterns vary within East Asian communities as well. Chinese immigration to the United States began in the 1850s and was concentrated primarily in the West. Chinese immigration during this time consisted mainly of male laborers who mined for gold, worked in agriculture, or worked for railroad companies (Batalova & Greene, 2025). Other American laborers working these positions began to blame Chinese immigrants, who had less bargaining power and were willing to accept lower wages, for driving down labor costs (U.S. Foreign Relations - Office of the Historian, n.d.). This contributed to the developing resentment and hostility towards Chinese laborers, which ultimately led to the passing of the Chinese Exclusion Act in 1882. This policy heavily restricted Chinese immigration until World War II, which saw its repeal. Compared with the first waves of Chinese immigration to the United States, immigration following the 1965 Immigration Reform Act consisted of a professional, highly skilled class, including students and businessmen (Batalova & Greene, 2025). Today, Chinese Americans have a median household income of $84,215, with 56% holding a bachelor's degree or higher and 43.2% reporting limited English proficiency (AAPI Data, 2022; see Table II).

*Korean*

Korean immigration to the United States began in the early 1900s and saw its increase following both World War II and the Korean War (Yoon I-j, 1997). The first major wave of Korean immigration to the United States began in 1903, with many Koreans immigrating to Hawaii to work on sugar plantations (Shin, 2024). The next wave of immigration followed the Korean War and consisted primarily of the wives of U.S troops in South Korea, children orphaned by the war and adopted by American families, and a professional class that included students and businessmen (Chung, n.d.). Korean immigration to the U.S. increased further in 1965, with the shift to a quota-based immigration system. This final wave of immigration was the largest, and prioritized skilled professionals and family reunification (Shin, 2024). This wave included immigrants from white-collar backgrounds from a variety of occupations that were moving to the United States by choice, seeking economic opportunity (Korean American Foundation - Greater Washington, n.d.). Korean Americans now report a median household income of $74,958 and high levels of educational attainment, with 59% holding a bachelor's degree or higher (AAPI Data, 2022; see Table II).

*Japanese*

Similarly to Korean and Filipino immigration, Japanese immigration to the United States started at sugar plantations in Hawaii (Stanford Medicine, 2014). The first Japanese immigrants arrived in Hawaii, and then on the West Coast of the mainland United States, particularly in California. The Meiji Restoration in 1868 and the rapid move towards industrialization created political and economic unrest in Japan that acted as a push factor for many Japanese immigrants (Library of Congress, n.d.). Most of the Japanese immigrants arriving at this time were young men seeking economic mobility, as students, male laborers, or both (Akiba, 2006). The Chinese Exclusion Act of 1882 led to a high demand for labor in positions such as agriculture and railways that Japanese immigrants began to fill (Minidoka National Historic Site, 2019). They inherited the racial resentment and hostility that White Americans had developed against Chinese immigrants as a result. Between 1908 and 1924, the next wave of Japanese immigration began. Japanese women began immigrating to the United States as "picture brides" of Japanese American men, a result of a loophole in the Gentleman's Agreement, which limited the immigration of Japanese men, but allowed the wives of immigrants already in the United States to join them (Akiba, 2006). It was very difficult for Japanese American men to start a family, as many could not afford to return home to search for a wife, and interracial marriage between Japanese men and White women was forbidden by law. Picture brides resulted, which were arranged marriages between Japanese American men and Japanese women, where the prospective couple would be introduced to each other via photographs. Following World War II, another nearly 45,000 Japanese women immigrated to the United States as war brides of U.S. military men. After 1965, Japanese immigration to the United States prioritized family reunification and skilled labor, though decreasing drastically in number compared to Japanese immigration between 1900 and 1920 (Karthick, 2015). Today, the median household income for Japanese Americans is $84,861, with 52% receiving a bachelor's degree or higher (AAPI Data, 2022; see Table II).

| Group | Median Household Income | Bachelor's or Higher (%) | Limited English Proficiency (%) | Immigration Reasons | |
|---|---|---|---|---|---|
| | | | | Employment | Family |
| Asian Indian | $125,319 | 75% | 18.2% | 54% | 43% |
| Filipino | $97,816 | 50% | 20.5% | 20% | 78% |
| Asian (Overall) | $91,828 | 55% | 31.9% | — | — |
| Japanese | $84,861 | 52% | 22.4% | 46% | 51% |
| Chinese | $84,215 | 56% | 43.2% | 34% | 54% |
| Korean | $74,958 | 59% | 38.8% | 66% | 34% |
| Vietnamese | $71,014 | 32% | 48.9% | 6% | 93% |
| NHPI | $66,406 | 19% | 12.2% | — | — |
| Other Asian | $65,793 | 46% | 27.7% | — | — |

**Table II. This table was compiled using data from AAPI's "State of Asian (Americans), Native Hawaiians, and Pacific Islanders in the United States" Report from June of 2022 (AAPI Data, 2022).** Note: Native Hawaiians and Pacific Islanders (NHPI).

As Table II illustrates, Asian American and Pacific Islander groups differ in socioeconomic and educational outcomes. While Indian Americans report the highest median household income and bachelor's degree attainment, Native Hawaiians and Pacific Islanders (NHPI) report the lowest levels of educational attainment and earnings. Vietnamese Americans also report relatively low levels of educational attainment alongside high rates of limited English proficiency. These wide disparities directly counter the model minority myth by demonstrating that the "Asian" category is far from uniform; instead, socioeconomic status and educational outcomes vary significantly across subgroups. Treating Asian (Americans) as a monolithic high-achieving group obscures the challenges faced by many communities within the broader category.

The disparities reinforce calls within STEM education research for disaggregating Asian (American) experiences. For example, Zhang et al (2025) challenge the assumption that Asian (Americans) students are overrepresented in STEM fields. During interviews, students revealed that multiple aspects of their identity impacted their identity formation in physics. Their findings suggested that intersections regarding gender, socioeconomic background, and national origin may make the representation of Asian (Americans) in STEM more complex. These multiple, intersecting identities are often disguised by the broader model minority stereotype.

## Conceptual Framework

This study draws on two Critical Race Theory (CRT) informed frameworks: (1) Asian Critical Theory (AsianCrit) (Museus & Itikar, 2013)and (2) Quantitative Critical Race Theory (QuantCrit) (Castillo & Strunk, 2024), which together guide the study's conceptual framing and analytic decisions. Rather than applying CRT directly, the analysis centers AsianCrit to foreground the racialization of Asian students in physics education and to challenge homogenizing narratives such as the model minority myth that flattens racial and ethnic differences. QuantCrit extends these commitments into the quantitative domain by showing that racial categories, data aggregation, and modeling choices can reproduce or conceal inequities. By integrating AsianCrit and QuantCrit, the study aligns its conceptual commitments with its methodological approach: AsianCrit identifies the racialized processes that shape Asian students' experiences, and QuantCrit ensures that these insights inform decisions about data disaggregation, model specification, and interpretation of findings.

### 1. Asian Critical (AsianCrit) theory

AsianCrit builds on Critical Race Theory (CRT) by addressing the distinct racialization processes and sociohistorical positioning of Asian communities in the United States. While Museus and Itikar (2013) outlined seven tenets of AsianCrit, our analysis centers on three that are most relevant to our study:

a. Asianization recognizes how dominant U.S. society imposes homogenous and stereotypical identities onto Asian communities, flattening their cultural and ethnic differences. For instance, Asian students are often labeled as the "model minority," a stereotype that can obscure academic challenges and discourage them from seeking support.
b. Strategic (anti)essentialism emphasizes recognizing both shared and divergent experiences among Asian subgroups, resisting the notion of a monolithic Asian identity. Disaggregated educational data, for example, can highlight disparities between Southeast Asian and East Asian students, informing more targeted and equitable interventions.
c. Intersectionality identifies how intersections between social identities, such as race, gender, class, and language, shape complex educational experiences. In this paper, we focused on the intersection of different ethnic and racialized identities within the Asian community (e.g., White-Korean, etc.). For example, Asian Indian and Chinese American groups, who report among the highest rates of bachelor's degree attainment, may also draw upon insider knowledge and social capital that facilitate persistence in STEM fields, resources that institutions restrict for other minoritized groups (Seymour & Hewitt, 1997).

2. Quantitative Critical Race Theory (QuantCrit)

Gillborn et al. (2018) describe QuantCrit as a critical response to address racism and biases embedded in quantitative research methods. Quantitative methods in fields such as education and psychology were influenced by the eugenics movement, which promoted racial hierarchies and White supremacy (Castillo & Strunk, 2024). QuantCrit combines Critical Race Theory with quantitative methods to challenge these oppressive histories and address current inequalities (Gillborn et al., 2018).

QuantCrit also questions the common belief that quantitative research is objective and neutral, instead recognizing that all data and methods contain inherent biases. The goal of QuantCrit is to use a racial equity perspective to critically evaluate quantitative research and promote fairness and social justice in education (Anonymous et al., 2022). Our work focuses on two tenets that align with our research question, based on the definition in (Castillo & Strunk, 2024):

a. Numbers are Not Neutral. QuantCrit asserts that numbers and quantitative methods are not neutral because they reflect human choices influenced by societal biases and dominant cultural norms. Our study examines conventional racial groupings in education to highlight hidden inequities masked by aggregation.
b. Categories are Neither Natural Nor Inherent: QuantCrit recognizes racial categories as socially constructed and context-dependent. In this study, we examine disaggregated data on Asian identities (e.g., Asian Indian, Chinese, White-Korean, and Filipino) to reveal differences that aggregated categories (e.g., URM and non-URM) often obscure.

3. Theory-informed methods

Guided by AsianCrit and QuantCrit, we selected quantitative methods aligned with our conceptual framework and research questions. AsianCrit informed our decision to challenge the model minority myth by exploring the heterogeneity of Asian (American) students across ethnic/racial groups. QuantCrit shaped our methodological approach by leading us to question the use of aggregated racial categories and to instead examine inequities through disaggregated subgroup analyses. We applied MAIHDA (Evans et al., 2024), which enabled us to model outcomes across subgroups of Asian (Americans) students in introductory physics.

# Methods

## Data collection and cleaning

We used the data from the LASSO platform (Anonymous, 2018), an online assessment system that supports the administration, scoring, and analysis of research-based assessments (RBAs). LASSO provides anonymous data from consenting students for research. The dataset contained social identity information (i.e., races, ethnicities) and test performance from Fall 2015 to Fall 2023. The dataset included 16,810 students collected from 493 calculus-based physics courses across 64 institutions. We selected data from two common physics RBAs: (1) Force

Concept Inventory - FCI (Hestenes et al., 1992) with 10,723 students; and (2) Force and Motion Conceptual Evaluation - FMCE (Thornton & Sokoloff, 1998) with 6,087 students (see Table III). The FCI and FMCE assess students' knowledge of Newtonian mechanics.

| | N | Pretest (%) | Posttest (%) | Pre and Posttest (%) |
|---|---|---|---|---|
| FCI | 10,723 | 8,591 (80%) | 7,555 (70%) | 5,423 (51%) |
| FMCE | 6,087 | 4,888 (80%) | 3,963 (65%) | 2,764 (45%) |
| Overall | 16,810 | 13,479 (80%) | 11,518 (69%) | 8,187 (49%) |

**Table III. Number and percentage of student responses for the FCI and FMCE assessments.** The N column represents the number of students who completed at least 1 administration (pre- or post). The other columns indicate the number of students who completed the pretest, posttest, or both, along with the percentage relative to the total.

We filtered the dataset to include only responses from calculus-based courses that administered the FCI or FMCE. We then removed the pretest or posttest score if the student took less than 5 minutes on the assessment or answered less than 80% of the test items. Among students who provided a pretest, posttest, or both, 20% were missing the pretest, and 31% were missing the posttest.

## Data imputation

To handle missing data, we employed Blimp (Du et al., 2022; Enders & Little, 2022; Keller, 2021; J. Kim et al., 2026). We specified an imputation model with a hierarchical structure in which we treated student-level variables as Level 1, course-level variables as Level 2, and institution-level variables as Level 3 (students nested within courses, and courses within institutions). We imputed 25 times, including course-ID and institution-ID as cluster variables to reflect this three-level structure.

In our analysis, we identified unique social groups based on race information from the LASSO survey. Among 366 distinct social identity groups, 144 were Asian (e.g., Asian Indian, Chinese, Filipino, Korean, Japanese, Vietnamese, etc). To align with our research question, we retained the label "pan-Asian strata" for the aggregated category collected from Fall 2015 to Spring 2022. For the disaggregated data collected from Summer 2022 through Fall 2023, we anonymized the Asian subgroups and relabeled them as Group 1, Group 2, Group 3, and so forth. This approach directs attention to comparative patterns across groups rather than to the performance of any particular subgroup. Across the 144 groups, we focused on the pan-Asian strata and the 19 subgroups that each had at least 10 student responses on either the pretest or posttest (see Figure 1 and Table VIII in the Appendix for sample sizes).

## Data analysis

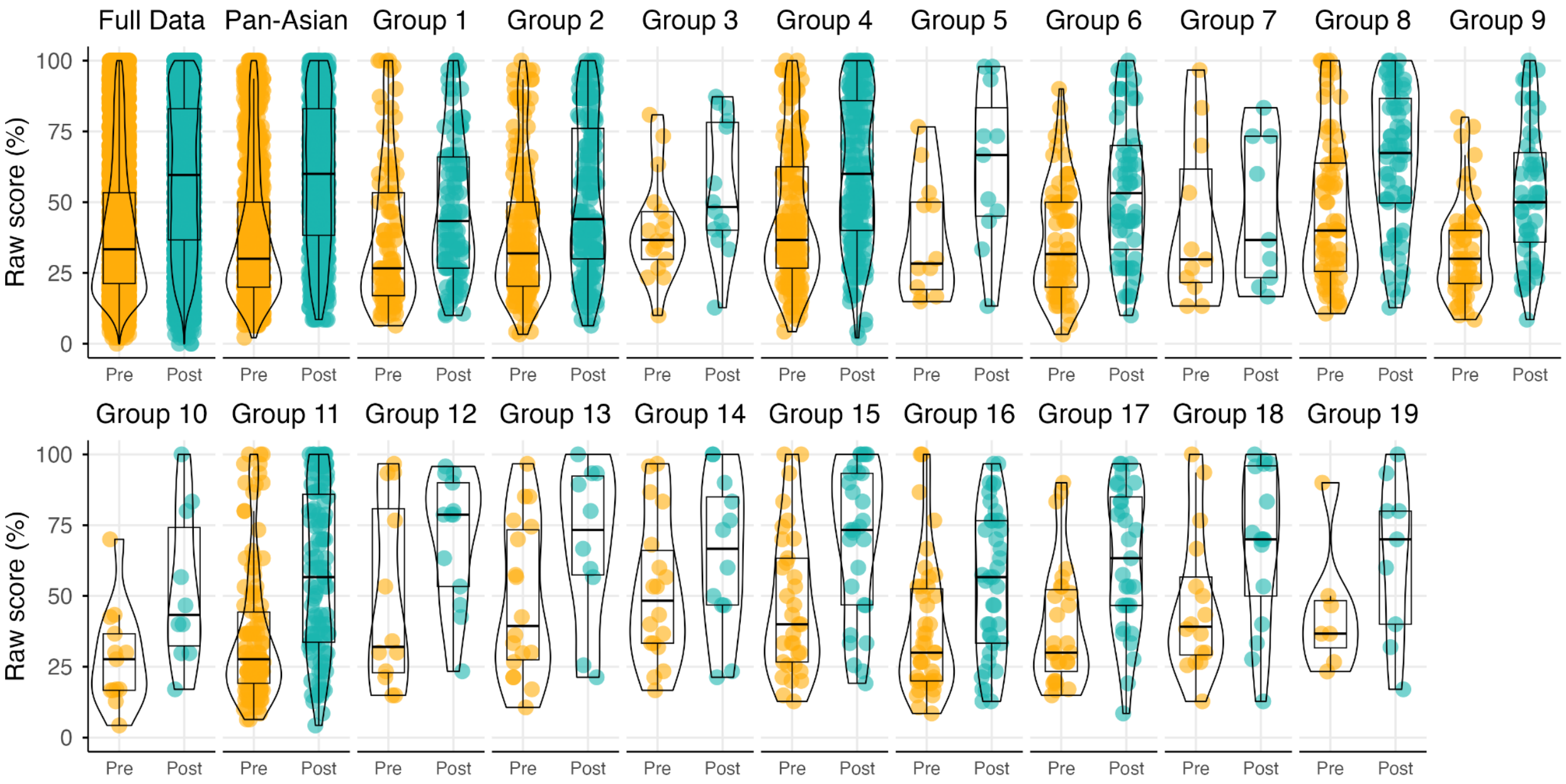

**Figure 1. The distribution of raw scores (%) for the full dataset, pan-Asian strata and the 19 Asian subgroups in pretest and posttest.** *Note: We collected data on the pan-Asian strata prior to 2022, and afterward, we collected data on Asian subgroups (Groups 1-19).*

In educational research, MAIHDA (Evans et al., 2024) is a statistical approach used to analyze group-level characteristics (e.g., students, social identity groups, courses, schools, or classrooms) to explain individual educational outcomes, such as academic performance (Evans et al., 2024; Anonymous et al., 2024). Unlike some quantitative intersectional models, MAIHDA does not employ multiple interaction terms to capture intersecting social identities (e.g., ethnicity, gender, socioeconomic status). Instead, it nests individuals within strata (i.e., intersectional identity groups defined by combinations of social identities). By introducing a group-level error term, MAIHDA captures the unique outcomes for each identity combination, simplifying model complexity while addressing intersectionality (Evans et al., 2024). MAIHDA then adds this strata-level error term to the fixed effects when predicting a social strata's outcome. We identified the social strata based on the students' race and ethnicity (see Figure 2 - Strata).

To improve model predictions and better represent social identity interactions, we used effect coding (0.5 for present, −0.5 for absent) for racial indicator variables (Castillo & Kistler, 2026; Mayhew & Simonoff, 2015). We included terms for whether students were retaking the class, which assessment they took, and races/ethnicities, which defined the pan-Asian strata, and the 19 Asian subgroups.

We implemented the Bayesian MAIHDA model using the *brms* package (Bürkner, 2017) with the multiple imputed datasets. The MAIHDA model was a three-level cross-classified multilevel model nesting tests (level 1) within students (level 2), within courses (level 3a), and within strata (level 3b) (see Figure 2 for the cross-classified MAIHDA). The analysis used the *brm_multiple* function (Bürkner, 2017) with 2,000 iterations per chain, a 1,000 warm-up, and a 1,000 sampling across four chains to ensure stable and accurate posterior estimation. Because the model drew from 25 imputed datasets, it generated 100,000 posterior samples. We then combined the fixed effects and random effects from the MAIHDA model to generate posterior distributions for each of the 20 strata, the aggregated pan-Asian strata, and the 19 disaggregated Asian subgroups.

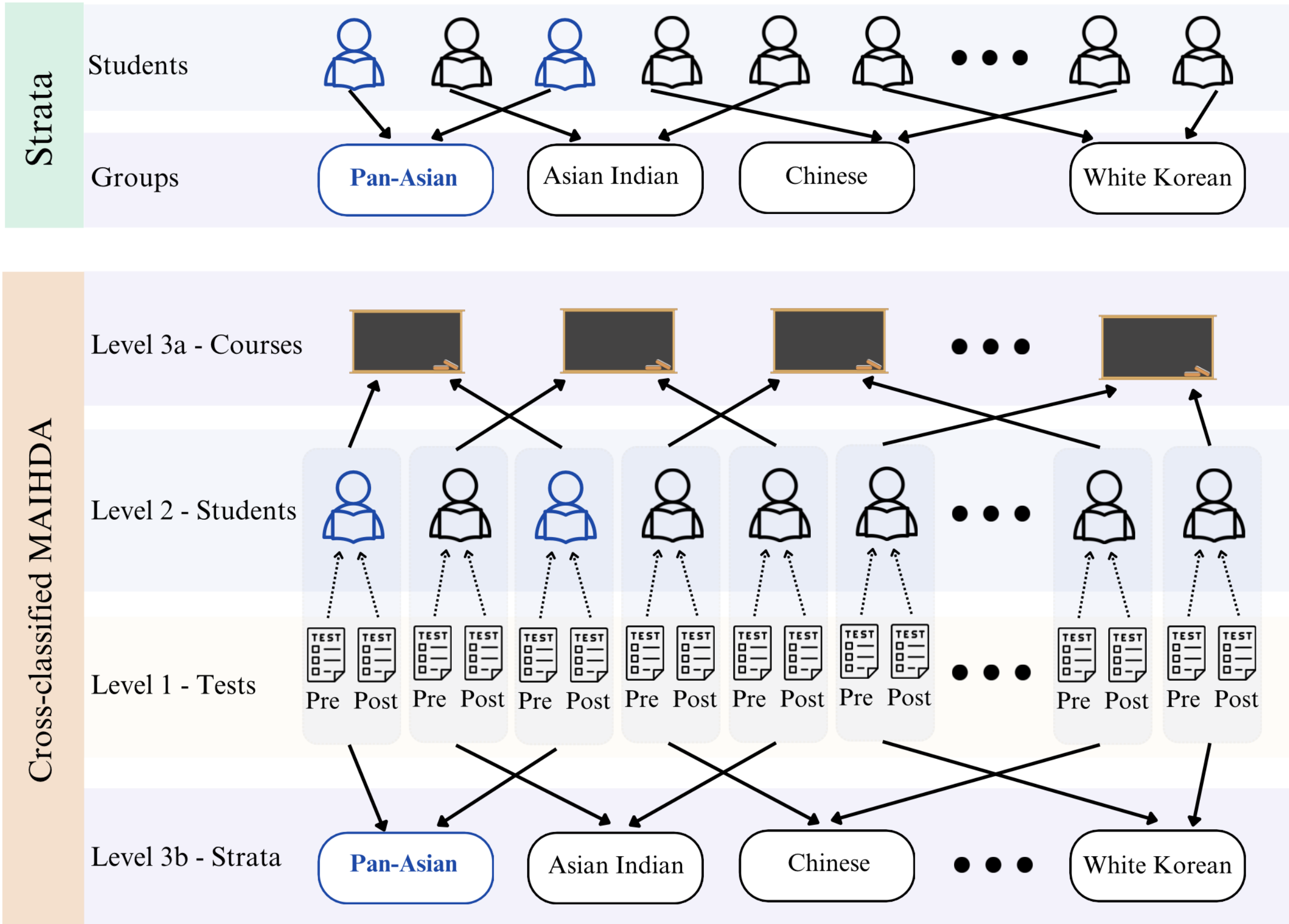

**Figure 2. The MAIHDA model.** Strata illustrates how students are nested within (multi-) racial/ethnic strata (e.g., Asian, Asian Indian, Chinese, White, Korean). Cross-classified MAIHDA depicts the three-level model: Level 1 contains pre- and posttest scores, Level 2 nests students, and Level 3 cross-classifies students by course enrollment (3a) and racial/ethnic strata (3b). *Note: We collected data on the pan-Asian strata prior to 2022, and afterward we collected Asian subgroups (e.g., Asian Indian, Chinese, and White Korean).*

## Quantifying and interpreting the heterogeneity across Asian Subgroups

We answered the research question with two complementary analyses of the posterior distributions from the MAIHDA model. First, we computed posterior distributions of differences by subtracting the pan-Asian strata's predicted mean from each subgroup's predicted mean for every posterior draw, then summarized each difference using the posterior mean, the 95% credible interval (CI) of the difference, and the posterior probability that the subgroup mean was higher or lower than the pan-Asian strata. Second, we used the mean absolute error (MAE) to quantify the overall heterogeneity of the 19 Asian subgroups relative to the pan-Asian strata. Together, the posterior differences provide the subgroup-level detail needed to interpret the MAE, which serves as the primary, reproducible summary of heterogeneity.

From the posterior distributions, we identified the spread of predicted means across all 20 groups by noting the highest and lowest values. We examined the degree of overlap in the 95% CI for the groups with the highest and lowest predicted means to characterize uncertainty in those extreme differences.

For each subgroup, we summarized the posterior distribution of differences using three quantities: (1) the posterior mean difference, (2) the 95% CI of the difference, and (3) the posterior probability that the subgroup mean was higher or lower than the pan-Asian strata. We report the posterior mean as the point summary because the posterior distributions were approximately symmetric; readers may treat it as equivalent to the posterior median in this context. In the results, we report that the number of subgroups had 95% CI that excluded zero and were higher than the pan-Asian strata. A full summary table for each subgroup at pretest and posttest appears in the Appendix (see Table VI).

To interpret the magnitude of observed differences, we used the pan-Asian strata's pre-to-post gain as a benchmark. This gain reflects the average improvement of Asian students over a 16-week course, including its associated uncertainty captured in the posterior distribution. We expressed the range between the highest and lowest subgroup predicted means as a proportion of this benchmark gain, allowing us to interpret subgroup differences on the scale of a semester of instruction while acknowledging that both the subgroup range and the benchmark gain carry posterior uncertainty.

We summarized the overall variability across subgroups using the MAE. For each of the 100,000 posterior samples, we calculated the absolute difference between each of the 19 subgroup scores and the pan-Asian strata score, then averaged these 19 differences into a single averaged absolute error. We repeated this across all posterior samples and summarized the distribution using the posterior mean and 95% CI of the averaged absolute error. A larger MAE indicates greater heterogeneity across subgroups.

### Variance Partition Coefficient (VPC)

The VPC is the proportion of the total variance in the outcome that lies between strata:

$$\text{VPC} = \sigma^2_u \,/\, (\, \sigma^2_u + \sigma^2_e \,)$$

It ranges from 0 to 1 (commonly reported as a percentage). Between-strata variance ($\sigma^2_u$) the spread of the strata means around the grand mean (intercept). In the null model it captures the total variation attributable to the strata, before adjusting for any factor. Within-strata variance ($\sigma^2_e$) the residual variation among units that belong to the same strata. Statistically it is the intra-class correlation: the expected correlation between two units drawn at random from the same stratum, equivalently the share of total variation that the stratum structure accounts for. A VPC near 1 means stratum membership is highly informative about a unit's outcome - units within a stratum are very similar and differ sharply from those in other strata. A VPC near 0 means the strata are effectively interchangeable and membership carries little information about the outcome.

Proportional Change in Variance (PCV)

To separate additive from interaction effects, the null model is compared with an additive main-effects model. In additive main-effects model the additive contributions of the factors are absorbed into the fixed part, so the between-strata variance ($\sigma^2_u$) now measures only the variation that *remains* between strata after additive adjustment, that is, the portion not described by a consistent, additive combination of factor levels. This residual between-strata variation is what is attributable to interactions among the factors.

The PCV is the proportional reduction in between-strata variance from null model to additive main-effects model:

$$\text{PCV} = (\ \sigma^2_u\,(\text{null}) - \sigma^2_u\,(\text{additive})\ ) \,/\, \sigma^2_u\,(\text{null})$$

Reported as a percentage, the PCV is the share of the total between-strata variance explained by the additive main effects. Its complement,

$$1 - \text{PCV}$$

is the share left unexplained by additive effects and therefore attributable to *interaction* effects. A PCV close to 100% (with a near-zero residual VPC in additive main-effects model) indicates the between-strata structure is well described additively; a PCV meaningfully below 100% indicates that interactions among the factors are needed to reproduce the observed strata means. For binary outcomes the PCV is computed identically on the logit scale.

We fit four models in sequence to guide model selection using VPC and PCV. All models shared the same cross-classified three-level structure described above: pre- and posttest scores (level 1) nested within students (level 2), with students cross-classified by course (level 3a) and racial/ethnic strata (level 3b). The models differed in which predictors they included at the fixed-effects level.

- Model 1 was an intercept-only model with no random effects, used to establish the total unconditional variance as the denominator for VPC calculations in subsequent models; it is not reported separately.
- Model 2 added the full random-effects structure: random intercepts for course, student, and strata, plus a strata-by-test random slope allowing pre- and posttest differences to vary across strata. This model had no fixed predictors beyond the intercept, so its between-strata variance reflects the raw, unadjusted heterogeneity across groups.
- Model 3 retained that random-effects structure and added fixed effects for test timing (pretest vs. posttest), retake status, and assessment instrument (FCI or FMCE). These predictors adjust for differences that are not about race but that may vary across strata and account for some of the observed between-strata variance.

- Model 4 extended Model 3 by adding fixed effects for each racial/ethnic identifier (e.g., Asian, Chinese, Vietnamese, Japanese), producing the fully specified model aligned with the research question.

The VPC from Model 2 quantified the total between-strata variance before covariate adjustment. The PCV from Model 2 to Model 3 and from Model 2 to Model 4 showed how much of that variance each set of predictors explained. Tables IV and V report the VPC and PCV results for all models, here VPC values are posterior means with 95% CI. PCV values compare each model to Model 2 (the baseline nesting model).

| Model | Component | VPC | 95% CI |
|---|---|---|---|
| Model 2: nesting + strata | Course | 10.3 | [8.7, 12.2] |
| | Strata | 4.2 | [1.3, 9.9] |
| Model 3: non-race predictors | Course | 8.2 | [6.9, 9.6] |
| | Strata | 1.7 | [0.8, 2.9] |
| Model 4: fully specified | Course | 8.3 | [7.0, 9.7] |
| | Strata | 0.7 | [0.3, 1.5] |

**Table IV: Variance Partition Coefficients (VPC) across the four-model sequence.** VPC values represent the posterior mean percentage of total outcome variance attributable to each model component, with 95% credible intervals (CI).

| Model | Component | PCV | 95% CI |
|---|---|---|---|
| Model 2 → 3 (academic predictors) | Course | 35.1 | [18.9, 48.9] |
| | Strata | 57.7 | [−11.7, 88.7] |
| Model 2 → 4 (all predictors) | Course | 35.4 | [19.3, 49.0] |
| | Strata | 81.6 | [48.1, 95.7] |

**Table V: Proportional Change in Variance (PCV) for course and strata components relative to Model 2.** PCV values represent the posterior mean percentage reduction in between-component variance when predictors are added, with 95% credible intervals (CI). Positive values indicate variance reduction; negative lower bounds reflect posterior uncertainty in small strata variance estimates.

In Model 2, strata-level variance accounted for 4.2% (95% CI [1.3%, 9.9%]) of the total outcome variance, indicating that meaningful differences in physics scores existed between racial/ethnic groups before any predictors were introduced. Adding academic and assessment-level predictors in Model 3 reduced that between-strata variance by 57.7% relative to Model 2 (PCV: 95% CI [−11.7%, 88.7%]), with the strata VPC dropping to 1.7% (95% CI [0.8%, 2.9%]). This reduction indicates that test timing, retake status, and instrument type explained a substantial portion of why groups differed. Adding race/ethnicity fixed effects in Model 4 reduced between-strata variance by 81.6% relative to Model 2 (PCV: 95% CI [48.1%, 95.7%]), with the strata VPC dropping further to 0.7% (95% CI [0.3%, 1.5%]). These two PCV values are both computed relative to Model 2 as the shared baseline and are not additive. The residual 0.7% between-strata variance in Model 4 represents the interaction patterns among identities, strata-level differences not captured by the additive fixed effects, which the strata random effects in Model 4 correctly model.

We selected Model 4 as the final model for reporting subgroup heterogeneity. Conceptually, including the race/ethnicity fixed effects aligns with the research question: to examine variation in physics knowledge across Asian racial/ethnic subgroups, the model must account for the additive contributions of those identities to produce well-calibrated strata-level predictions. Statistically, Model 4 produced the largest reduction in between-strata variance (PCV = 81.6% relative to Model 2), with the residual strata VPC of 0.7% indicating that nearly all of the systematic group differences were captured by the fixed effects. The remaining strata-level random effects in Model 4 therefore represent genuine interaction patterns, variance among strata arising from combinations of identities.

## Findings

Figure 3 shows the distribution of predicted pre- and posttest for the pan-Asian strata and the 19 Asian subgroups, based on Model 4 (the fully specified model selected based on VPC and PCV results), see Tables IV and V. Model 4 reduced between-strata variance by 81.6% relative to the baseline nesting model, with the residual strata VPC of 0.7% reflecting interaction patterns among identities captured by the strata random effects. We reported detailed predicted means and 95% CIs for each group in the Appendix (Table VI). While we include the model's coefficients in the Appendix (Table VII), we do not examine them directly because meaningful predictions require the combination of multiple coefficients. As we focus on differences between the pan-Asian strata and Asian subgroups, we do not make direct comparisons among specific subgroups to avoid ranking them and reinforcing negative stereotypes. Informing the heterogeneity in outcomes does not require ranking subgroups; it requires identifying the extent to which the aggregated pan-Asian strata obscures meaningful variation.

The pan-Asian strata began introductory calculus-based physics with a mean pretest score of 43.2 (95% CI [41.4, 44.9]) and ended with a mean posttest score of 57.6 (95% CI [55.8, 59.3]), a gain of 14.4 percentage points over the 16-week course. However, the disaggregated subgroup estimates revealed substantial heterogeneity that the aggregated pan-Asian strata obscured. At pretest, predicted means across the 19 Asian subgroups ranged from 35.1 (Group 9) to 50.9 (Group 18), a range of 15.8 percentage points. At posttest, predicted means ranged from 49.5 (Group 9) to 64.9 (Group 18), a range of 15.4 percentage points. In both administrations, the range in predicted means exceeded the pan-Asian strata's entire gain during a 16-week course. The 95% CI of the lowest- and highest-performing subgroups did not overlap with each other at either time point, pretest ([29.4, 40.7] vs. [45.1, 56.5]) or posttest ([43.9, 55.3] vs. [59.2, 70.6]). The posterior probability that Group 9's mean was lower than the pan-Asian strata exceeded 0.99 at both pretest and posttest; the posterior probability that Group 18's mean was higher than the pan-Asian strata likewise exceeded 0.99 at both administrations, providing strong evidence that these extreme subgroups differ from the aggregated group.

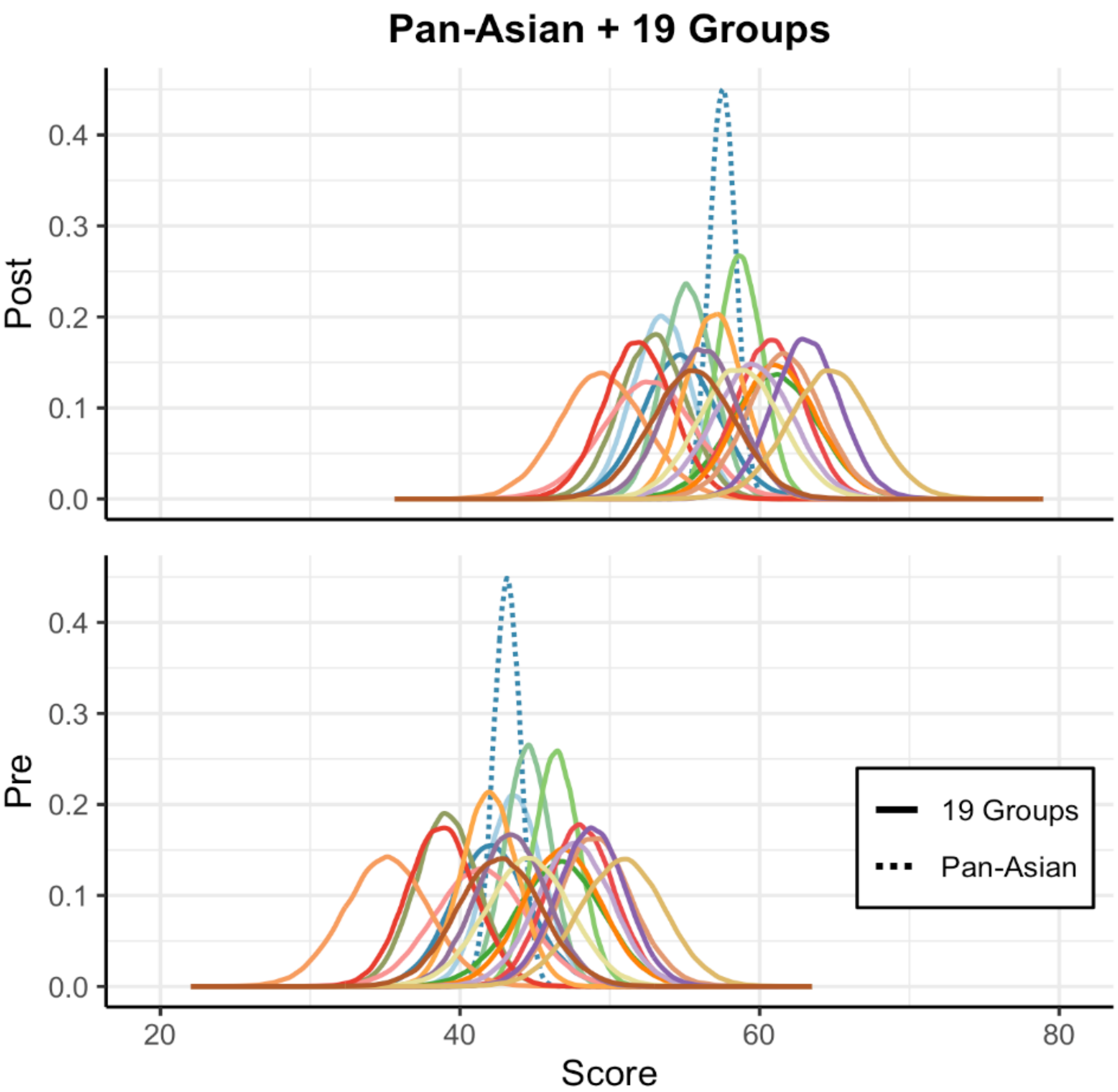

**Figure 3. Bayesian posterior distributions of pretest and posttest outcomes across the pan-Asian strata and 19 Asian subgroups.** *Note: We collected data on the pan-Asian strata prior to 2022, and afterward, we collected data on Asian subgroups (Groups 1-19).*

Across all 19 subgroups, we computed the posterior probability that each subgroup's mean differed from the pan-Asian strata. At pretest, 11 subgroups had posterior probabilities exceeding 0.95 that their mean differed from the pan-Asian strata; at posttest, this increased to 15 subgroups. The 95% CI of the difference between each of these subgroups and the pan-Asian strata excluded zero, indicating that the differences are unlikely to reflect sampling variability alone. Full posterior difference summaries for all 19 subgroups at pretest and posttest appear in Table VIII of the Appendix.

Heterogeneity across Asian subgroups was substantial. The mean absolute error between the 19 subgroup predicted means and the pan-Asian strata was 3.9 percentage points (95% CI [3.2, 4.6]) at pretest and 4.0 percentage points (95% CI [3.3, 4.7]) at posttest. Expressed relative to the pan-Asian strata's 16-week gain of 14.4 percentage points, the MAE corresponds to approximately 4–5 weeks of learning, on average, separating Asian subgroup performance from the pan-Asian aggregate.

## Discussion

Our findings reveal meaningful heterogeneity in predicted physics performance across Asian subgroups, challenging the assumption of uniform academic outcomes that aggregated analyses imply. From an AsianCrit perspective, this variation refutes the model minority myth by undermining the notion of a monolithic Asian identity and surfacing substantial differences in physics performance levels across subgroups at both the start and end of the course.

Across the 19 subgroups, performance differences were large and persisted from pretest to posttest. At pretest, predicted means spanned 15.8 percentage points, reflecting substantial variation in students' prior preparation in physics conceptual knowledge. At posttest, the spread remained similarly large at 15.4 percentage points, indicating that instruction did not substantially reduce pre-existing disparities. The relative ordering of subgroups remained largely consistent across administrations, suggesting that the course did not differentially benefit lower-performing subgroups.

To contextualize the magnitude of these differences, we interpret subgroup variation relative to the pan-Asian strata's 16-week learning gain, consistent with approaches used in educational research to express effect sizes in terms of instructional time (Kraft, 2020). The 15.8-point pretest spread and 15.4-point posttest spread each exceeded the pan-Asian strata's full course gain of 14.4 points, underscoring the scale of subgroup heterogeneity. The lowest-performing subgroup's posttest mean (49.5, 95% CI [43.9, 55.3]) approached the highest-performing subgroup's pretest mean (50.9, 95% CI [45.1, 56.5]). After a full semester of instruction, the lowest-performing subgroup ended the course at roughly the same level where the highest-performing subgroup began, indicating a gap of approximately a full course of learning separating these two groups. The MAE of approximately 4.0 percentage points at both time points, equivalent to 4–5 weeks of learning, represents a gap large enough to influence decisions about instructional support and resource allocation.

The posterior probability results reinforce this pattern. At pretest, 11 subgroups had greater than 95% posterior probability that their mean differed from the pan-Asian strata; at posttest, this increased to 15 subgroups. The two subgroups at the performance extremes (Group 9 and Group 18) each exceeded 0.99 posterior probability of differing from the pan-Asian strata at both time points, providing the strongest evidence that aggregation masks meaningful differences for these groups.

These findings underscore the importance of data disaggregation in educational research. Without subgroup-level analyses, inequities in performance and learning opportunities remain hidden. Disaggregated data enable researchers and educators to identify groups that may require additional support (Sbeglia & Nehm, 2024), better understand differences in educational experiences, and design more targeted and equitable interventions.

## Limitations

Our sample size limited the ability to represent all Asian groups and the intersectional aspects of their social identities we would like to include in this analysis (e.g., gender and first-generation college status). These factors vary across groups: for example, only 6.2% of Japanese students in our data identified as first-generation college students, compared to over 80% of Asian Hispanic/Latino students (see Table IX). In the institutional contexts, the proportion of students attending R1 universities varied from 17.6% to 63.6% across the 19 subgroups (see Table X). Prior research demonstrates that these variables shape academic outcomes: first-generation college students often face systemic barriers to success (Jehangir, 2010), institutional type can shape access to resources and support (Kuh, 2010), and gender dynamics influence performance and retention, particularly in STEM fields (Ong et al., 2011). Future research should extend intersectional analyses, such as exploring interactions between gender and Asian ethnicity identities, to further examine the sources and structures of educational disparities.

## Conclusions

Aggregated data can produce a misleading statistical picture of a heterogeneous racial group whose subgroups experience different patterns in educational outcomes (Teranishi et al., 2020). This problem is especially acute when aggregation masks disparities in opportunities and achievement among certain subgroups. By using MAIHDA within AsianCrit and QuantCrit frameworks, this study revealed large disparities in outcomes across Asian subgroups - disparities that the pan-Asian strata concealed. Predicted means represent the central tendency of each subgroup’s performance, not the absolute performance of any individual member. Readers should interpret subgroup differences as differences in group averages, recognizing that individuals within each subgroup vary around those means.

The data cannot identify the specific causes of the disparities observed, but prior research provides important context for interpreting them. Migration histories, English language proficiency, and access to educational opportunities vary across Asian subgroups and shape academic outcomes (AAPI Data, 2022). Understanding how these factors contribute to physics outcomes across Asian subgroups remains an important direction for future research.

Disaggregated data support identifying disparities at a more granular level, enabling researchers and educators to direct resources where they are most needed and to address the specific needs of particular student groups. Increased disaggregation also introduces methodological challenges, including smaller sample sizes, wider credible intervals, and reduced interpretability. Researchers must balance granularity with statistical precision, guided by theoretical frameworks such as QuantCrit, research questions, and analytic goals.

Collecting identity data in fine-grained, flexible ways enables researchers to aggregate or disaggregate groups as appropriate to the sample size and analytical context. The LASSO platform’s identity options made this level of analysis possible, and MAIHDA provided a powerful approach for examining subgroup-level heterogeneity. Fine-grained identity data collection can support combining data across studies to advance equity-focused, intersectional analyses in physics education and beyond.

# Appendix

| Group | N | Pretest | | Posttest | |
|---|---|---|---|---|---|
| | | M | 95% CI | M | 95% CI |
| Pan-Asian strata | 1134 | 43.2 | 41.4 - 44.9 | 57.6 | 55.8 - 59.3 |
| 1 | 134 | 43.6 | 39.9 - 47.7 | 53.5 | 49.6 - 57.3 |
| 2 | 259 | 44.8 | 41.8 - 47.7 | 55.1 | 51.8 - 58.5 |
| 3 | 20 | 42.4 | 37.3 - 47.7 | 55 | 49.9 - 60.3 |
| 4 | 245 | 46.5 | 43.4 - 49.6 | 58.7 | 55.8 - 61.6 |
| 5 | 13 | 46.7 | 40.7 - 52.5 | 61.2 | 55.3 - 67.0 |
| 6 | 82 | 39.2 | 35.0 - 43.3 | 53.1 | 48.7 - 57.4 |
| 7 | 16 | 41.4 | 35.3 - 47.5 | 52.6 | 46.3 - 58.7 |
| 8 | 91 | 48 | 43.6 - 52.4 | 60.8 | 56.3 - 65.2 |
| 9 | 68 | 35.1 | 29.4 - 40.7 | 49.5 | 43.9 - 55.3 |
| 10 | 14 | 38.7 | 34.3 - 43.2 | 51.9 | 47.4 - 56.5 |
| 11 | 153 | 42 | 38.4 - 45.6 | 57 | 53.2 - 60.8 |
| 12 | 15 | 46.9 | 41.6 - 52.3 | 61.4 | 56.1 - 67.0 |
| 13 | 19 | 49.3 | 44.5 - 54.4 | 62 | 56.9 - 67.2 |
| 14 | 21 | 47.6 | 42.5 - 52.7 | 59.5 | 53.9 - 64.8 |
| 15 | 39 | 48.6 | 44.1 - 53.1 | 63.1 | 58.5 - 67.6 |
| 16 | 54 | 43.4 | 38.7 - 48.1 | 56.1 | 51.4 - 60.9 |
| 17 | 35 | 44.5 | 38.8 - 50.1 | 58.7 | 53.3 - 64.2 |
| 18 | 20 | 50.9 | 45.1 - 56.5 | 64.9 | 59.2 - 70.6 |
| 19 | 11 | 42.6 | 37.0 - 48.3 | 55.6 | 50.0 - 61.2 |
| Average of 19 subgroup strata | | 44.3 | 39.6 - 48.9 | 57.4 | 52.7 - 62.1 |

**Table VI. Predicted mean scores across the pan-Asian strata and 19 subgroups on the pre- and posttest.** The table reports the number of student responses (*N*), predicted mean scores (*M*), and 95% compatibility interval (CI) for each group. Predicted mean scores varied by 15.8 percentage points across groups on the pretest [35.1-50.9] and by 15.4 percentage points on the posttest [49.5-64.9]. *Note: We collected data on the pan-Asian strata prior to 2022, and afterward we collected data on Asian subgroups (Groups 1-19).*

| Coefficient | Estimate | Est. Error | 95% Compatibility Interval | |
|---|---|---|---|---|
| | | | Min | Max |
| Intercept | 53.2 | 3.3 | 46.8 | 59.6 |
| *Fixed effects* | | | | |
| Pan-Asian strata | 4.9 | 1.6 | 1.9 | 8.1 |
| Race/Ethnicity A | 4.9 | 1.5 | 1.9 | 7.8 |
| Race/Ethnicity B | -1.5 | 2.0 | -5.4 | 2.5 |
| Race/Ethnicity C | 3.3 | 1.8 | -0.3 | 6.7 |
| Race/Ethnicity D | 2.8 | 2.0 | -1.2 | 6.6 |
| Race/Ethnicity E | 7.7 | 2.3 | 3.3 | 12.2 |
| Race/Ethnicity F | 3.1 | 1.9 | -0.6 | 6.8 |
| Race/Ethnicity G | 4.0 | 0.8 | 2.4 | 5.5 |
| Race/Ethnicity H | -0.6 | 1.8 | -4.1 | 2.9 |
| Race/Ethnicity I | 1.1 | 2.5 | -3.7 | 5.9 |
| Race/Ethnicity J | -2.7 | 1.1 | -4.9 | -0.5 |
| Posttest | 13.2 | 0.7 | 11.9 | 14.6 |
| Strata | | | | |
| Pan-Asian strata - Pre | -1.5 | 1.5 | -4.5 | 1.4 |
| Group 1 (Race/Ethnicity F) - Pre | 0.8 | 1.7 | -2.7 | 4.3 |
| Group 2 (Race/Ethnicity C) - Pre | 1.8 | 1.6 | -1.3 | 5.2 |
| Group 3 (Race/Ethnicity J) - Pre | 0.5 | 2.0 | -3.5 | 4.5 |
| Group 4 (Race/Ethnicity A) - Pre | 1.9 | 1.6 | -1.2 | 5.3 |
| Group 5 (Race/Ethnicity A & D) - Pre | -0.6 | 2.0 | -4.7 | 3.4 |
| Group 6 (Race/Ethnicity H) - Pre | 0.0 | 1.8 | -3.5 | 3.5 |
| Group 7 (Race/Ethnicity I) - Pre | 0.5 | 2.0 | -3.5 | 4.7 |
| Group 8 (Race/Ethnicity E) - Pre | 0.5 | 1.8 | -3.0 | 4.2 |
| Group 9 (Race/Ethnicity B & J) - Pre | 0.5 | 1.8 | -3.1 | 4.1 |
| Group 10 (Race/Ethnicity D) - Pre | -0.4 | 2.0 | -4.6 | 3.6 |
| Group 11 (Race/Ethnicity A & D) - Pre | -0.4 | 1.7 | -3.7 | 2.9 |
| Group 12 (Race/Ethnicity G & F) - Pre | 0.1 | 2.0 | -3.9 | 4.2 |

| Group 13 (Race/Ethnicity G) - Pre | 0.8 | 2.0 | -3.1 | 4.9 |
|---|---|---|---|---|
| Group 14 (Race/Ethnicity G & C) - Pre | 0.7 | 2.0 | -3.2 | 4.8 |
| Group 15 (Race/Ethnicity G & A) - Pre | 0.1 | 1.9 | -3.6 | 4.0 |
| Group 16 (Race/Ethnicity G & H) - Pre | 0.3 | 1.9 | -3.4 | 4.1 |
| Group 17 (Race/Ethnicity G & I) - Pre | -0.3 | 2.0 | -4.2 | 3.6 |
| Group 18 (Race/Ethnicity G & J) - Pre | -0.5 | 2.0 | -4.6 | 3.4 |
| Group 19 (Race/Ethnicity G & B) - Pre | 0.4 | 2.1 | -3.6 | 4.7 |
| Pan-Asian strata - Post | -0.3 | 1.5 | -3.3 | 2.6 |
| Group 1 (Race/Ethnicity F) - Post | -2.6 | 1.8 | -6.4 | 0.8 |
| Group 2 (Race/Ethnicity C) - Post | -1.1 | 1.6 | -4.3 | 2.1 |
| Group 3 (Race/Ethnicity J) - Post | -0.1 | 2.0 | -4.1 | 3.9 |
| Group 4 (Race/Ethnicity A) - Post | 0.9 | 1.6 | -2.1 | 4.1 |
| Group 5 (Race/Ethnicity A & D) - Post | 0.6 | 2.1 | -3.4 | 4.8 |
| Group 6 (Race/Ethnicity H) - Post | 0.7 | 1.7 | -2.7 | 4.2 |
| Group 7 (Race/Ethnicity I) - Post | -1.5 | 2.1 | -5.9 | 2.5 |
| Group 8 (Race/Ethnicity E) - Post | 0.1 | 1.8 | -3.4 | 3.7 |
| Group 9 (Race/Ethnicity B & J) - Post | 0.4 | 1.8 | -3.2 | 4.1 |
| Group 10 (Race/Ethnicity D) - Post | 0.8 | 2.0 | -3.2 | 4.9 |
| Group 11 (Race/Ethnicity A & D) - Post | 1.3 | 1.7 | -2.0 | 4.8 |
| Group 12 (Race/Ethnicity G & F) - Post | 1.4 | 2.0 | -2.5 | 5.7 |
| Group 13 (Race/Ethnicity G) - Post | 0.2 | 2.0 | -3.9 | 4.2 |
| Group 14 (Race/Ethnicity G & C) - Post | -0.7 | 2.0 | -4.9 | 3.3 |
| Group 15 (Race/Ethnicity G & A) - Post | 1.3 | 1.9 | -2.4 | 5.3 |
| Group 16 (Race/Ethnicity G & H) - Post | -0.2 | 1.9 | -3.9 | 3.6 |
| Group 17 (Race/Ethnicity G & I) - Post | 0.7 | 1.9 | -3.0 | 4.6 |
| Group 18 (Race/Ethnicity G & J) - Post | 0.3 | 2.0 | -3.7 | 4.3 |
| Group 19 (Race/Ethnicity G & B) - Post | 0.2 | 2.1 | -3.9 | 4.3 |

**Table VII. Model coefficients, estimated error, and 95% compatibility intervals.** Race/ethnicity categories are relabeled as A, B, C, and so forth and the intersectional social identity groups as the anonymized subgroup reporting (Groups 1-19). *Note: We collected data*

*on the pan-Asian strata prior to 2022, and afterward we collected data on Asian subgroups (Groups 1-19).*

| **Group** | **Number of responses** | | |
|---|---|---|---|
| | **Total** | **Pretest** | **Posttest** |
| Pan-Asian strata | 1134 | 997 | 677 |
| 1 | 134 | 102 | 109 |
| 2 | 259 | 197 | 187 |
| 3 | 20 | 16 | 16 |
| 4 | 245 | 191 | 196 |
| 5 | 13 | 12 | 11 |
| 6 | 82 | 71 | 63 |
| 7 | 16 | 11 | 9 |
| 8 | 91 | 75 | 69 |
| 9 | 68 | 54 | 52 |
| 10 | 14 | 13 | 10 |
| 11 | 153 | 117 | 113 |
| 12 | 15 | 12 | 13 |
| 13 | 19 | 18 | 10 |
| 14 | 21 | 18 | 12 |
| 15 | 39 | 33 | 29 |
| 16 | 54 | 42 | 41 |
| 17 | 35 | 27 | 31 |
| 18 | 20 | 17 | 14 |
| 19 | 11 | 7 | 9 |

**Table VIII: Information of pan-Asian strata and 19 subgroups with cutting off at least 10 responses in pre- or posttest.** *Note: We collected data on the pan-Asian strata prior to 2022, and afterward we collected data on Asian subgroups (Groups 1-19).*

| Group | N | Gender (%) | | | | First Gen (%) |
|---|---|---|---|---|---|---|
| | | Man | Woman | Non Binary/ Transgender | Missing Data | |
| Pan-Asian strata | 1134 | 64.5 | 34.9 | 0.4 | 0.2 | 38.8 |
| 1 | 134 | 70.1 | 28.4 | 1.5 | 0 | 42.5 |
| 2 | 259 | 63.3 | 36.3 | 0.4 | 0 | 22.6 |
| 3 | 20 | 70 | 30 | 0 | 0 | 82.5 |
| 4 | 245 | 63.3 | 35.9 | 0.8 | 0 | 41.6 |
| 5 | 13 | 84.6 | 15.4 | 0 | 0 | 53.8 |
| 6 | 82 | 70.7 | 25.6 | 3.7 | 0 | 18.3 |
| 7 | 16 | 56.2 | 37.5 | 6.2 | 0 | 6.2 |
| 8 | 91 | 57.1 | 42.9 | 0 | 0 | 11 |
| 9 | 68 | 64.7 | 33.8 | 1.5 | 0 | 42.6 |
| 10 | 14 | 57.1 | 42.9 | 0 | 0 | 46.4 |
| 11 | 153 | 64.1 | 33.3 | 2 | 0.7 | 47.1 |
| 12 | 15 | 46.7 | 40 | 6.7 | 6.7 | 6.7 |
| 13 | 19 | 73.7 | 26.3 | 0 | 0 | 28.9 |
| 14 | 21 | 47.6 | 47.6 | 0 | 4.8 | 7.1 |
| 15 | 39 | 61.5 | 30.8 | 7.7 | 0 | 5.1 |
| 16 | 54 | 61.1 | 35.2 | 1.9 | 1.9 | 24.1 |
| 17 | 35 | 51.4 | 48.6 | 0 | 0 | 11.4 |
| 18 | 20 | 60 | 40 | 0 | 0 | 30 |
| 19 | 11 | 3.6 | 36.4 | 0 | 0 | 9.1 |
| Total | 2443 | | | | | |

**Table IX: Descriptive statistics for gender and first generation across pan-Asian strata and 19 subgroups**. *Note: We collected data on the pan-Asian strata prior to 2022, and afterward we collected data on Asian subgroups (Groups 1-19).*

| Group | Number of responses | Number of institutes | R1 (%) | R2 (%) | Other (%) |
|---|---|---|---|---|---|
| Pan-Asian strata | 1134 | 30 | 31.2 | 26.8 | 42 |
| 1 | 134 | 29 | 41.8 | 33.6 | 24.6 |
| 2 | 259 | 34 | 51 | 28.6 | 20.5 |
| 3 | 20 | 8 | 25 | 25 | 50 |
| 4 | 245 | 33 | 54.7 | 18.4 | 26.9 |
| 5 | 13 | 11 | 38.5 | 30.8 | 30.8 |
| 6 | 82 | 20 | 23.2 | 50 | 26.8 |
| 7 | 16 | 10 | 56.2 | 25 | 18.8 |
| 8 | 91 | 22 | 39.6 | 11 | 49.5 |
| 9 | 68 | 16 | 17.6 | 23.5 | 58.8 |
| 10 | 14 | 10 | 21.4 | 28.6 | 50 |
| 11 | 153 | 28 | 30.1 | 27.5 | 42.5 |
| 12 | 15 | 11 | 53.3 | 13.3 | 33.3 |
| 13 | 19 | 9 | 36.8 | 26.3 | 36.8 |
| 14 | 21 | 12 | 57.1 | 23.8 | 19 |
| 15 | 39 | 20 | 48.7 | 23.1 | 28.2 |
| 16 | 54 | 23 | 40.7 | 25.9 | 33.3 |
| 17 | 35 | 18 | 45.7 | 25.7 | 28.6 |
| 18 | 20 | 12 | 50 | 25 | 25 |
| 19 | 11 | 7 | 63.6 | 18.2 | 18.2 |

**Table X: Descriptive statistics for institution types by students and institutions across pan-Asian strata and 19 subgroups.** *Note: We collected data on the pan-Asian strata prior to 2022, and afterward we collected data on Asian subgroups (Groups 1-19).*